\documentclass[12pt]{article}
\font\sdopp=msbm10
\def\CI {\sdopp {\hbox{C}}}
\title{
A weaker geodesic completeness\\
 and 
Clifton-Pohl torus}
\author{Claudio Meneghini\\
{\small\tt 
clamen@dimat.unipv.it}
}
\begin{document}
\maketitle
\bibliographystyle{plain} 
\parindent=8pt
\font\cir=wncyb10
\def\Iu{\cir\hbox{YU}}
\def\Ze{\cir\hbox{Z}}
\def\pe{\cir\hbox{P}}
\def\Ef{\cir\hbox{F}}
\font\sdopp=msbm10
\def\DI{\sdopp{\hbox{D}}}
\def\ESSE {\sdopp {\hbox{S}}}
\def\ERRE {\sdopp {\hbox{R}}}
\def\CI {\sdopp {\hbox{C}}}
\def\ENNE{\sdopp {\hbox{N}}}
\def\ZETA{\sdopp {\hbox{Z}}}
\def\PI {\sdopp {\hbox{P}}}
\def\M{\hbox{\boldmath{}$M$\unboldmath}} 
\def\N{\hbox{\boldmath{}$N$\unboldmath}} 
\def\P{\hbox{\boldmath{}$P$\unboldmath}} 
\def\Y{\hbox{\boldmath{}$Y$\unboldmath}} 
\def\tr{\hbox{\boldmath{}$tr$\unboldmath}} 
\def\f{\hbox{\boldmath{}$f$\unboldmath}} 
\def\u{\hbox{\boldmath{}$u$\unboldmath}} 
\def\v{\hbox{\boldmath{}$v$\unboldmath}} 
\def\U{\hbox{\boldmath{}$U$\unboldmath}} 
\def\V{\hbox{\boldmath{}$V$\unboldmath}} 
\def\W{\hbox{\boldmath{}$W$\unboldmath}} 
\def\id{\hbox{\boldmath{}$id$\unboldmath}} 
\def\alph{\hbox{\boldmath{}$\aleph$\unboldmath}} 
\def\bet{\hbox{\boldmath{}$\beta$\unboldmath}} 
\def\gam{\hbox{\boldmath{}$\gamma$\unboldmath}} 
\def\U{\mathop{u}\limits}
\def\f{\hbox{\boldmath{}$f$\unboldmath}} 
\def\g{\hbox{\boldmath{}$g$\unboldmath}} 
\def\h{\hbox{\boldmath{}$h$\unboldmath}} 
\def\IM{\hbox{\boldmath{}$i$\unboldmath}} 

\def\Ch{\hbox{\rm Ch}}
\def\CIRC{\mathop{\tt o}\limits}
\def\quan{\vrule height6pt width6pt depth0pt}
\def\QUAN{\ \quan}
\def\BETA{\mathop{\beta}\limits}
\def\GAMMA{\mathop{\gamma}\limits}
\def\VI{\mathop{v}\limits}
\def\UI{\mathop{u}\limits}

\def\oi{\mathop{\omega}\limits}
\def\ei{\mathop{\eta}\limits}
\def\FI{\mathop{\varphi}\limits}

\def\VII{\mathop{V}\limits}
\def\WI{\mathop{w}\limits}
\def\ZETA{\mathop{Z}\limits}
\def\BETA{\mathop{\beta}\limits}
\def\GAMMA{\mathop{\gamma}\limits}
\def\VI{\mathop{v}\limits}
\def\UI{\mathop{u}\limits}
\def\VII{\mathop{V}\limits}
\def\WI{\mathop{w}\limits}
\def\ZETA{\mathop{Z}\limits}
\def\ssqrt#1{\left(#1\right)^{1/2}}
\def\sssqrt#1{\left(#1\right)^{-1/2}}
\def\TTT{\sl}
\def\BBB{\sl}

\newtheorem{definition}{Definition}
\newtheorem{defi}[definition]{D\'efinition}
\newtheorem{lemma}[definition]{Lemma}
\newtheorem{lemme}[definition]{Lemme}
\newtheorem{proposition}[definition]{Proposition}
\newtheorem{theorem}[definition]{Theorem}        
\newtheorem{theoreme}[definition]{Th\'eor\`eme}        
\newtheorem{corollary}[definition]{Corollary}  
\newtheorem{corollaire}[definition]{Corollaire}  
\newtheorem{remark}[definition]{Remark}  
\newtheorem{remarque}[definition]{Remarque}
  
\font\sdopp=msbm10
\def\ERRE {\sdopp {\hbox{R}}}
\def\QU {\sdopp {\hbox{Q}}}
\def\CI {\sdopp {\hbox{C}}}
\def\DI {\sdopp {\hbox{D}}}
\def\ENNE{\sdopp {\hbox{N}}}
\def\ZETA{\sdopp {\hbox{Z}}}
\def\PI {\sdopp {\hbox{P}}}

\def\M{\hbox{\tt\large M}}
\def\N{\hbox{\tt\large N}}
\def\T{\hbox{\tt\large T}}

\def\P{\hbox{\boldmath{}$P$\unboldmath}} 
\def\tr{\hbox{\boldmath{}$tr$\unboldmath}} 
\def\f{\hbox{\large\tt f}} 
\def\g{\hbox{\tt\large g}}

\def\F{\hbox{\boldmath{}$F$\unboldmath}} 
\def\G{\hbox{\boldmath{}$G$\unboldmath}} 
\def\L{\hbox{\boldmath{}$L$\unboldmath}} 
\def\h{\hbox{\boldmath{}$h$\unboldmath}} 
\def\e{\hbox{\boldmath{}$e$\unboldmath}} 

\def\u{\hbox{\boldmath{}$u$\unboldmath}} 
\def\v{\hbox{\boldmath{}$v$\unboldmath}} 
\def\U{\hbox{\boldmath{}$U$\unboldmath}} 
\def\V{\hbox{\boldmath{}$V$\unboldmath}} 
\def\W{\hbox{\boldmath{}$W$\unboldmath}} 
\def\id{\hbox{\boldmath{}$id$\unboldmath}} 
\def\alph{\hbox{\boldmath{}$\alpha$\unboldmath}} 
\def\bet{\hbox{\boldmath{}$\beta$\unboldmath}} 
\def\gam{\hbox{\boldmath{}$\gamma$\unboldmath}} 
\def\pphi{\hbox{\boldmath{}$\varphi$\unboldmath}} 
\def\ppsi{\hbox{\boldmath{}$\psi$\unboldmath}} 
\def\Ppsi{\hbox{\boldmath{}$\Psi$\unboldmath}} 
\def\labelle #1{\label{#1}}
\begin{abstract}
We propose a new definition of geodesic
completeness, based on analytical continuation
in the complex domain: we apply this idea to
Clifton-Pohl torus, relating, for each geodesic,
completeness to the value of a function of initial conditions, called 'impulse'.
\end{abstract}
\section{Foreword}
We propose a weaker definition of
geodesic completeness and use it to
classify geodesics of 'Clifton-Pohl
torus' $\T$ (a compact, geodesically incomplete, Lorentz manifold, see 
\cite{oneill}, 7.16).
We need the idea
of a {\it holomorphic metric} on a complex manifold $\M$ (see \cite{lebrun}
): it amounts to a nondegenerating symmetric section 
of
the twice covariant holomorphic tensor bundle 
 ${\cal T}_{0}^{2}\M$.
Of course, it carries no 'signature';
however, by simmetry, it induces
a canonical Levi-Civita's connexion on $\M$, allowing
geodesics to be defined as 
auto-parallel paths.
Moreover, if $\M$ arises as a 'complexification' of a semi-Riemannian manifold $\N$,
it is easily seen that the real geodesics  of $\N$ are restrictions to the real axis of the complex ones of $\M$ and vice versa (see \cite{lebrun}).
This fact allows us to 'flank' isolated singularities on the real line by running along complex trips, i.e. to 'connect'
geodesics which, in the usual sense,
are completely unrelated.

We suggest an idea
of 
our notion of {\sl completeness}
(see also definition \ref{completessa})
:
given a complexification $d:\N\rightarrow\M$
and a real analytic curve $\gamma:[a,b]\rightarrow\N$,
$\gamma$ will be 
told to be 
{\sl complete}
provided that 
$d\circ\gamma$ can be continued to all points
in the real line,  with at most
a discrete set of exceptional values,
taking 'real values' (i.e. in $d(\N)$).
\section{Basic definitions and lemmata}
In the following,  ${\cal U}$ will be a region in the complex plane and $\M$
a complex manifold: the idea of the analytical continuation of a holomorphic mapping element
(or of a germ) 
$f:{\cal U}\rightarrow\M$  is well known and amounts to a quadruple $Q_{\M}=(S,\pi,j,F)$, where
$S$ is a connected Riemann surface over a region of $\CI$,
$\pi\,\colon\, S\rightarrow \CI$ is a nonconstant holomorphic mapping 
such that $U\subset \pi(S)$,
$j\,\colon\, U\rightarrow S$ is a holomorphic  immersion such that $\pi\circ j=id\vert_{U}$
and 
$F\,\colon\, S\rightarrow \M$ is a holomorphic mapping such that $F\circ j=f$.
Each finite branch point is 
kept into account by the fact of lying 'under' some critical point of $\pi$; it is a well known
result that there exists a unique maximal analytical continuation,
called the {\BBB Riemann surface}, of $\left({\cal U},f   \right)$.
\begin{lemma}
Let $\f$ and $\g$ be two
$\CI$-valued
 holomorphic germs each one inverse of the
 other, admitting the representatives
$({\cal U},f  )$, resp. $({\cal V},g  )$
;
 let $(   R,\pi,j,F)$ and $(S,\rho,\ell,G )$
be their respective Riemann surfaces: then $F(R)=\rho(S)$.
\labelle{inverse}
\end{lemma}
{\bf Proof: a)} 
 $F(R)\subset\rho(S)$: let $\xi\in R$ and $F(\xi)=\eta$; there exist:
a neighbourhood ${\cal U}_1$ of $\xi$;
open subsets ${\cal U}_2\subset\pi({\cal U}_1   )$ and ${\cal V}_2\subset F({\cal U}_1   )$ and 
a biholomorphic function $g_2:{\cal V}_2\rightarrow{\cal U}_2$, with inverse
$f_2:{\cal U}_2\rightarrow{\cal V}_2$
{such that}:
$({\cal U}_2,f_2   )$ and $({\cal U},f  )$ are connectible and so are
$({\cal V}_2,g_2   )$ and $({\cal V},g  )$.
There exist two holomorphic immersions 
$\widetilde j:{\cal U}_2\rightarrow R\hbox{ and }
\widetilde\ell:{\cal V}_2\rightarrow S$
such that $\pi\circ\widetilde j=\id$ and $\rho\circ\widetilde\ell=\id$.
Let ${\cal V}_1=F(U)_1$ and 
$
\Sigma=\{(x,y   )\in{\cal U}_1\times{\cal V}_2: F(x)=y\}
$;
 let $J:{\cal V}_2\rightarrow\Sigma$
be defined by $ J(v)=(\widetilde j\circ  g_2(v),v )$.
Then $ (\Sigma,pr_2,J,\pi\circ pr_1   )$ is an analytical continuation of $({\cal V}_2,g_2   )$; indeed $ \pi\circ pr_1\circ J=\pi\circ\widetilde j\circ g_2=g_2$. But $({\cal V}_2,g_2   )$ is connectible with $({\cal V},g     )$, so
$ (\Sigma,pr_2,J,\pi\circ pr_1   )$ is an analytical continuation of $({\cal V},g )$.
Eventually, $\exists h\in{\cal O}(\Sigma,S)$
such that $\rho\circ h=pr_2$: hence 
$
\eta=pr_2(\xi,\eta   )=\rho\circ h
(\xi,\eta   )\in\rho(S   )
$.

{\bf b)} $\rho(S)\subset F(R)$: let $s\in S$: there is a neighbourhood  $V$ of $s$ in $S$ such that $V\setminus\{s\}$ consists of regular points both of $\rho$ and $G$.
This means that for each $s^{\prime}\in V\setminus\{s\}$ there exists a holomorphic function element $({\cal V}^{\prime},
\widetilde g_{s^{\prime}})$ (with 
${\rho(s^{\prime})},
\in{\cal V}^{\prime}   $) connectible with $({\cal V},g   )$ and a holomorphic immersion $\widetilde\ell:{\cal V}^{\prime}\rightarrow V$.
By a) already proved, $G(s)\in\pi(R)$, hence
$\exists p\in R$ such that $\pi(p)=G(s)$ and
a neighbourhood $W$ of $p$ in $R$ such that $ \pi^{-1}(\widetilde g({\cal V}^{\prime}   )   )\bigcap W\not=\emptyset$.
Set
$ W^{\prime}=\pi^{-1}(\widetilde g({\cal V}^{\prime}   )   )\bigcap W$:
suppose, without loss of generality, $\pi$ invertible on $W^{\prime}$: hence there exists a open holomorphic immersion $\widetilde j:\widetilde g({\cal V}^{\prime}   )\rightarrow W$.
Therefore, for each $\zeta\in\widetilde j(\widetilde g({\cal V}^{\prime}   )   )$, there exists $ \eta\in\widetilde\ell({\cal V}^{\prime}   )$ such that $ F(\zeta)=F(\widetilde j \circ \widetilde g\circ \rho(\eta)  )$.
Now, by definition of analytical continuation, $ F\circ\widetilde j\circ\widetilde g=\id$, hence 
$ F(\zeta)=\rho(\eta)$.
Consider now the holomorphic function 
$\Xi:W\times V\rightarrow\CI$ 
defined by setting $ \Xi(w,v   )=F(w)-\rho(v)$: we have
$\Xi\equiv 0$ throughout
$ {\widetilde j(\widetilde g({\cal V}^{\prime}   )   )\times\widetilde\ell({\cal V}^{\prime})}$,
hence on $W\times V$, which in turn implies $ F(p)=\rho(s)$: this eventually implies that $\rho(S)\subset F(R)$.
\QUAN

\begin{definition}
A real-analytic curve $\gamma$ in a 
real-analytic manifold $\N$ admitting a complexification $d:\N\rightarrow\M$ is 
{\sl complete}
provided that the Riemann surface
$(S,\pi,j,G)$ of $d\CIRC\gamma$ is such that
$\ERRE\setminus\pi(G^{-1}(d(\N)))$
is 
a discrete set
\labelle{completessa}
\end{definition}
\begin{definition}
{\TTT A 
holomorphic 
metric on} $\M$ is an everywhere maximum-rank 
symmetric section 
of
the twice covariant holomorphic tensor bundle 
 ${\cal T}_{0}^{2}\M$.
A {\BBB holomorphic 
Riemannian manifold} is a complex manifold endowed with a 
holomorphic 
metric.
\label{riemann}
\end{definition}

We report the existence-and-uniqueness theorem of o.d.e's theorey in the complex domain:
let $W_0
$ be a complex $N-$tuple,
$z_0\in\CI$;
let $F$ be a $\CI^N-$valued holomorphic mapping in 
$\prod_{j=1}^N\DI
(W_0^j,b    
)\times\DI
(z_0,a   
)$,
($a,b\in\ERRE$)
with $C^0-$norm $M$ and
$C^0-$norm of each
${\partial F}/{\partial w^j}$ ($j=1..N$) not exceeding $K\in\ERRE$.

\begin{theorem}
If  
$r<min(a,b/M,1/K)$,
$\exists\,!$ a holomorphic mapping 
$
W\colon \DI
(z_0,r    
)$ $\rightarrow
\prod_{j=1}^N\DI
(W_0^j,b    
)
$
such that 
$W^{\prime}=F(W(z),z)$ and
$W(z_0)=W_0$.
(see e.g. \cite{hille}, th 2.2.2, \cite{ince} p.281-284)
\end{theorem}
As a consequence, for each point $p$ in
a holomorphic Riemannian manifold and each holomorphic tangent vector $X$ at $p$, there exists a unique holomorphic geodesic starting at $p$ with velocity $X$.
 
\section{The Clifton-Pohl torus}
Consider now 
$\N:=\ERRE^2
\setminus \{0\}$, with the Lorentz metric
${du\odot dv}/({u^2+v^2})
$;
the group $D$ generated by scalar
multiplication by $2$ is a group of isometries of 
$\N$; 
its action is properly dicontinuous, hence
$\T=\N/D$ is a Lorentz surface.
Topologically, $\T$
is  the closed annulus $1\leq\varrho\leq 2$, with boundaries identified by the action of $D$, i.e. a torus; notwithstanding, $\T$ is geodesically incomplete, since
$t\mapsto\left(1/(1-t),0\right)$ is a geodesic of $\hbox{\tt M}$ (see
\cite{oneill}).
In the following, we shall study directly 
$\N$ rather than
$\T$, since our conclusions could be easily pushed down with respect to the action of 
$D$.
Consider now the holomorphic Riemannian manifold
$
\M=\left[\CI^2
\setminus ((1,i)\CI \cup (1,-i)\CI),
{du\odot dv}/({u^2+v^2})
   \right]
$.

\begin{lemma}
The geodesic equations of 
both $\M$ and $\N$ are:
$
\UI^{\bullet\bullet}=
{2u}/({u^2+v^2})
\UI^{\bullet}{}^2$,
$
\VI^{\bullet\bullet}=
{2v}/({u^2+v^2})
\VI^{\bullet}{}^2
$; they are meant to be real or complex
depending on the fact that they concern $\hbox{\tt M}$ or $\hbox{\tt N}$.
\labelle{geodeq}
\end{lemma}
\begin{proposition}
All null geodesics of $\N$ are complete.
\vskip0.1truecm
\noindent
{\bf Proof:}\rm\ we may deal with the only case
$v=const:=A$.
Lemma \ref{geodeq} imply
$\UI^{\bullet\bullet}=
{2u}/({u^2+A^2})
\UI^{\bullet}{}^2$, which is solved by 
$t\mapsto (C-Bt)^{-1}$ if $A=0$ and by 
$t\mapsto\tan(At+B)$ if $A\not=0$, for suitable real constants $B$ and $C$.
The above functions are restrictions of meromorphic functions, hence, by definition \ref{completessa}, yield complete
geodesics.
\QUAN
\end{proposition}
We turn to nonnull geodesics of $\N$:
\begin{lemma}
The Cauchy's problem 
$
\FI^{\bullet}=2A\,\Ch\varphi
\sqrt{B^2-2/A\,\Ch\varphi}$
$ \varphi(0)=\varphi_0$,
(with $B^2-2/A\,\Ch\varphi_0>0$)
has complete solutions, in the real domain, with respect to the canonical complexification, if and only if $0<AB^2\leq 2$.
\labelle{fit}
\end{lemma}
{\bf Proof:}
set 
$F(\varphi)=2A\,\Ch\varphi
\sqrt{B^2-2/A\,\Ch\varphi}$
and
$G(\varphi):=
\int_{\varphi_0}^{\varphi}
d\,\nu/F(\nu)
$, where by the integral sign 
we mean the choice of the only primitive of $1/F$ vanishing at $\varphi_0$.
Rewrite the problem in the form $G(\varphi)=
\id$: this shows that $\varphi$
and $G$ are inverse elements of holomorphic functions in neighbourhoods of $\varphi_0$
and $G(\varphi_0)$.

Suppose $AB^2\geq 2$ or $AB^2<0$: then 
$F$ never vanishes; since $1/F(\nu)=O(\e^{-\vert\nu\vert})$ as $\nu\to\infty$,
$G$ takes a bounded set of values, hence, by lemma \ref{inverse}, $\varphi$ is not complete.  

If, instead, $0<AB^2\leq 2$, then
there exists a branch of $F$ admitting a zero on the real line, hence there exists a branch $\tilde f$ of $1/F$ whose absolute value takes all large enough values.
However $\tilde f$ can be analytically continued, by admitting complex trips, up to 
$\{\varphi:\Ch\varphi\geq 2/AB^2\}$, in such a way that an even function $f$ is yielded.

Now $\vert\int_{\varphi_0}^{\varphi}f(\nu)\,d\nu\vert$ takes {\sf all} {positive} values; but $g:=\int_{\varphi_0}^{\varphi}f$ is an odd function plus a real constant
on $\{\varphi:\Ch\varphi\geq 2/AB^2\}$, hence it takes {\sf all} real values with at most the exception of its asympotical value $\sigma$.
Thus, if $(S,\varrho,\ell,H)$ is the Riemann surface of $\varphi$, then,
by lemma \ref{inverse},
$\varrho(H^{-1}(\ERRE))\cap\ERRE\supset g(\ERRE)\supset\ERRE
\setminus\{\sigma\}$.
\QUAN

\begin{definition}
The {\sf impulse function} 
$\P:T\N\setminus\{{\tt null\  vectors}\}\rightarrow\ERRE$ is defined by setting $\P(\alpha,\beta,x,y)=
(\alpha^2+\beta^2)^{-1}
(2\alpha\beta+\alpha^2y/x+\beta x/y)$.
\end{definition}

\begin{theorem}
A nonnull geodesic $\gamma$ starting
from $(\alpha,\beta)$, with velocity
$(x,y)$ is complete if and only
if $0<\P(\alpha,\beta,x,y)\leq 2$.
\labelle{principal}
\end{theorem}
{\bf Proof:}
we may suppose
$\alpha\not=0$ and
$\beta\not=0$.
Moreover, we have $x\not=0$ and
$y\not=0$.
The equations in lemma \ref{geodeq}
can be integrated once to yield:
\begin{equation}
\UI^{\bullet}\VI^{\bullet}=
A(u^2+v^2),\quad u/\UI^{\bullet}+v/\VI^{\bullet}=B
,
\labelle{intprimm}
\end{equation}
where $A=xy/(\alpha^2+\beta^2)$
and $B=\alpha/x+\beta/y$; note that $AB^2=\P(\alpha,\beta,x,y)$.

Introduce now the supplementary hypothesis that $u>0$ and $v>0$:
by performing the change of coordinates $u=\e^{\omega}$,
$v=\e^{\eta}$, (\ref{intprimm})
is turned into
\begin{equation}
\oi^{\bullet}\ei^{\bullet}=
2A\, \Ch(\omega-\eta),\quad
1/\oi^{\bullet}+1/\ei^{\bullet}=B
.
\labelle{intnoeuv}
\end{equation}
We can solve with respect to 
$\oi^{\bullet}$ and $\ei^{\bullet}$, getting
\begin{equation}
\cases
{
\oi^{\bullet}=
2\left(B-\sqrt{B^2-2/[A\, \Ch(\oi-\ei)]}
   \right)^{-1}\cr
\ei^{\bullet}=
2\left(B+\sqrt{B^2-2/[A\, \Ch(\oi-\ei)]}
   \right)^{-1}
}.
\labelle{intnoeuv2}
\end{equation}
Subtract and set $\varphi:=\oi-\ei$; this yields the equation in $\varphi$ studied in lemma \ref{fit},
with the appropriate initial value $\varphi(0)=\log(u/v)$; this Cauchy's problem has complete solutions
if and only if $0<\P(\alpha,\beta,x,y)\leq 2$.

Now the fact that $\varphi$ is incomplete easily implies that
so is $\gamma$.
Suppose, instead, that $\varphi$
is complete: from (\ref{intnoeuv2}),
we get
that both
$\ei^{\bullet}$ and $\oi^{\bullet}$ is complete; since
passing to a primitive preserves completeness,
so are $\ei$ and $\oi$: 
but $u=e^{\oi}$ and
$v=e^{\ei}$: this eventually implies that
$\gamma$ is complete.

To remove the hypothesis that $u>0$ and $v>0$, consider two geodesics $\gamma$, $\delta$, starting from, say, $(\alpha,0   )$, the former with velocity
$(x,y)$ and the latter $(x,-y)$ ($y>0$).
The first order systems, like (\ref
{intprimm}), of $\gamma$ and $\delta$ differ only in the signs of constants in their first equations.
Thus, the equations of those pieces of $\gamma$
lying in $Q_1=\{u>0,v>0\}$ and of those ones of $\delta$ lying in $Q_2=\{u>0,v<0\}$ are transformed into the same system
(\ref{intnoeuv}) by performing the change of coordinates
$(u,v)=(e^{\omega}, e^{\eta})$
in $Q_1$
, resp. $(u,v)=(e^{\omega}, -e^{\eta})$ in $Q_2$; an analogous argument holds for the other octants.
It is easily seen that if a nonnull geodesic intersects one of the coordinate axes at a point $P$, it does with finite (nonnull) velocity, hence it can be analytically continued across $P$, changing octant:
thus, once obtained the (maximal) curve 
$t\mapsto
(\omega(t),\eta(t))$, we can reconstruct 
the original (maximal) geodesic 
$t\mapsto
(u(t),v(t))$ 
by choosing the only smooth curve starting from
$(\alpha,\beta)$ whose graph is contained in the set
$(t,u,v\in\ERRE^3): u=\pm \e^{\omega(t)},v=\pm\e^{\eta(t)}$.
\QUAN

\end{document}